\documentclass[11pt,bibliography=totoc]{scrartcl}

\usepackage{soul}
\usepackage[utf8]{inputenc}
\usepackage[english]{babel}
\usepackage{graphicx,amsmath,amsfonts,amssymb,dcolumn,amsthm,slashed,bbm,xspace,pgffor,tikz,mathbbol,latexsym,cite,braket}
\usepackage{microtype}
\usepackage{mathtools}
\usepackage{nccmath}
\usepackage{lmodern}
\usepackage[normalem]{ulem}
\usepackage{xcolor}
\usepackage[colorlinks=true,citecolor=blue,urlcolor=blue,linkcolor=blue]{hyperref}
\usepackage[hang]{footmisc} 
\usepackage[affil-it]{authblk}
\usepackage{multicol}

\numberwithin{equation}{section}

\begin{document}

\title{\textbf{Confinement/Deconfinement temperature for a rotating quark-gluon plasma}}

\author{Nelson R.~F.~Braga$^{a}$\thanks{\href{mailto:braga@if.ufrj.br}{ braga@if.ufrj.br}} , ~ Luiz F.~Faulhaber$^{b}$\thanks{\href{mailto:luiz.ferreira@unifal-mg.edu.br}{ luiz.ferreira@unifal-mg.edu.br}}  ,~  Octavio C.~Junqueira$^a$\thanks{\href{mailto:octavioj@pos.if.ufrj.br}{octavioj@pos.if.ufrj.br}} }
\affil{\footnotesize $^{a}$ UFRJ --- Universidade Federal do Rio de Janeiro, Instituto de Física,\\
Caixa Postal 68528, Rio de Janeiro, Brasil}

\affil{\footnotesize $^{b}$ UNIFAL --- Instituto de Ciência e Tecnologia,\\ 
Universidade Federal de Alfenas, Caixa Postal 37715-400, \\ 
Poços de Caldas, MG, Brazil}

\date{}
\maketitle

\begin{abstract}

Non-central  heavy ion collisions lead to the production of a quark gluon plasma with angular momentum.  We investigate,  using holographic AdS/QCD models,  how does rotation of the medium affects the confinement/deconfinement  transition temperature $T_c$. 
In holographic models,  this transition is represented by a Hawking-Page process involving two 
asymptotically anti-de Sitter spaces.   The plasma is represented here  by extending  the holographic approach to anti-de Sitter spaces with cylindrical symmetry.  Then,  the rotation of the medium is introduced through a Lorentz boost.  We consider hard and soft wall AdS/QCD models.  In both cases we find it out that, as the rotational velocity $v$  increases, $T_c$ decreases,  following the expression $T_c(v) = T_c(0)/\gamma(v)$, where $\gamma(v)$ is the Lorentz factor.

\end{abstract}

\section{Introduction}
Heavy ion collisions, as produced in particle accelerators, lead to the formation  of hadronic matter in a state where quarks and gluons interact strongly but are not confined. This very short lived state of matter, called quark gluon plasma (QGP),  cannot  be observed directly. All the available information about QGP comes from   
  the particles reaching the detectors after hadronization.   For reviews about QGP, see for example \cite{Bass:1998vz,SCHERER1999279,Shuryak:2008eq,Casalderrey-Solana:2011dxg}.

  The QGP behaves approximately as in thermal equilibrium and there is a critical temperature $T_c$ below which the hadronic matter is confined. The value of $T_c$ is known to be affected by the density of the medium and also by the presence of magnetic fields $ \vec B $.  Non-central heavy ion collisions produce strong magnetic fields. For examples of studies of the effect of $ \vec B $  fields on $T_c$ see, for example, Refs.  \cite{Bali:2011qj,Fraga:2012fs,Ballon-Bayona:2013cta,Mamo:2015dea,Dudal:2015wfn,Li:2016gfn,Ballon-Bayona:2017dvv,Rodrigues:2017cha, Bohra:2019ebj, Bohra:2020qom}. 
  
  In  this type of collisions, besides the presence of strong magnetic fields, the plasma also has  a large angular momentum. The purpose of this article is to study the effect of rotation of the plasma on $T_c$. Holographic  AdS/QCD models, in particular  the hard wall \cite{Polchinski:2001tt,Boschi-Filho:2002wdj,Boschi-Filho:2002xih} and the soft wall  \cite{Karch:2006pv}  ones, provide an interesting tool to estimate $T_c$, as discussed in 
  \cite{Herzog:2006ra,BallonBayona:2007vp}. We extend the previous studies of confinement/deconfinement transition in these models to the case when the medium has angular momentum.

 This article is organized as follows: in section  \textbf{2}    we present  cylindrical AdS geometries and show how to implement a rotation in such systems.  Then in  section  \textbf{3}  we discuss the thermodynamics of the hard and soft wall models with cylindrical symmetry. In section \textbf{4}  we study the confinement/deconfinement transition for a rotating plasma dual to the AdS geometries. Finally in section \textbf{5}  we analyse our results and comment on them.

\section{Cylindrical black hole  in AdS space and rotation}

The holographic models we are going to consider are defined on planar 5-dimensional anti-de Sitter ($AdS_5$) geometries,
 that show up as solutions of Einsteins equations with negative cosmological constant $ \Lambda = - \frac{12}{L^2} $   and 
  constant curvature $R = - \frac{20}{L^2}  $.  At zero temperature the geometry is  an anti-de Sitter space with radius  $L$. At finite temperature there are two solutions. One is the thermal AdS space,  that in the Euclidean signature with compact time direction, is described by the metric
\begin{equation}\label{thermalAdS}
    ds^2= \frac{L^2}{z^2}\left( dt^2 + d\overrightarrow{x}^2 + dz^2 \right)\;.
\end{equation} 
The other is the AdS black hole space 
\begin{equation}\label{BHAdS}
ds^2= \frac{L^2}{z^2}\left( f(z) dt^2 + d\overrightarrow{x}^2 + \frac{dz^2}{f(z)} \right)\;,
\end{equation}
with $f(z) = 1 - z^4/z_h^4$,  where $z_h$ is the location of the horizon. 

For the black hole geometry, the time coordinate is periodic, with period $\beta$, related to the  temperature and the horizon position by $T = 1/\beta = 1/(\pi z_h)$ \cite{Hawking:1982dh}. In the thermal AdS case,  requiring that the assymtotic limits of the two geometries at $z=\epsilon$, with $ \epsilon \to 0 $,  are  the same, one finds that the period is $\beta^\prime = \pi z_h \sqrt{f(\epsilon)}$. 
  
\subsection{Cylindrical AdS geometries}

The geometry with cylindrical symmetry, relevant for studying a medium that is rotating, is obtained by considering the case when the boundary, spanned by the $x$ coordinates in the metrics \eqref{thermalAdS} and \eqref{BHAdS}, take, respectively,  the forms  
\begin{equation}\label{cthermalAdS}
	ds^2 = \frac{L^2}{z^2} \left(- dt^2 + l^2 d\phi^2 + \sum_{i=1}^2 dx_i^2 + dz^2\right)\;, 
\end{equation}
and
\begin{equation}\label{CBH}
	ds^2 = \frac{L^2}{z^2} \left(- f(z)dt^2 + l^2 d\phi^2 + \sum_{i=1}^2 dx_i^2 + \frac{dz^2}{f(z)}\right)\;, 
\end{equation}
where we now consider Lorentzian  signature,  $l$ is  the radius of a hyper-cylinder and $ 0 \le \phi \le 2 \pi $. 
 
Using  metrics \eqref{cthermalAdS} and \eqref{CBH} we will see that one can represent a medium that rotates with homogeneous velocity. This is clearly not the exact situation of a real QGP formed when two heavy ions collide. However this analysis will make it possible to understand qualitatively the effect of rotation on the value of the critical temperature $T_c$. This is similar to the approach followed in most studies about magnetic fields ${\vec B}$ acting on a plasma, as for example  in Refs.  \cite{Dudal:2014jfa,BRAGA2018186,BRAGA2019462,BRAGA2020135918,Braga:2021fey}, where uniform  magnetic fields are considered.

\subsection{Lorentz boost: Rotating cylindrical black hole}

 Rotation can be introduced by performing a coordinate transformation as is \cite{BravoGaete:2017dso, PhysRevD.97.024034}, corresponding to the change to an observer for which the angular coordinate is varying uniformly with time: 
\begin{eqnarray}
	t &\rightarrow& \frac{1}{\sqrt{1-l^2 \omega^2}} \left(t + l^2 \omega \phi \right)\;,\label{T1}\\
	\phi &\rightarrow& \frac{1}{\sqrt{1-l^2 \omega^2}} \left(\phi +  \omega t \right)\label{T2}\;,
\end{eqnarray}
where $\omega$ is the angular velocity of the rotating cylindrical black hole, associated with the rotation of the plasma via gauge/gravity duality. The resulting metric for the rotating BH reads \cite{Zhou:2021sdy}
\begin{eqnarray}\label{RotatingBHmetric}
	ds^2 &= & g_{tt} dt^2 + g_{t\phi}dt d\phi + g_{\phi t} d\phi dt + g_{\phi \phi} l^2 d\phi^2 \cr && + \,  g_{zz} dz^2 + g_{xx} \sum_{i=1}^2 dx_i^2\;,
\end{eqnarray}
with
\begin{eqnarray}
	g_{tt} &=& \frac{\gamma^2(\omega)L^2}{z^2}\left( \omega^2 l^2-f(z) \right)\;, \\
	g_{\phi \phi} &=& \frac{\gamma^2(\omega)L^2}{z^2}\left( 1-\omega^2 l^2f(z) \right)\;,\\
	g_{t \phi} &=& g_{\phi t} = \frac{\gamma^2(\omega)L^2}{z^2}\omega l^2 \left( 1-f(z) \right)\;,\\
	g_{zz} &=& \frac{L^2}{z^2 f(z)}\;,\\
	g_{xx} &=& \frac{L^2}{z^2}\;,\label{RotatingBH}
\end{eqnarray}
where $\gamma$ is the Lorentz factor,
\begin{equation}
	\gamma(\omega l ) = \frac{1}{\sqrt{1-l^2\omega^2}}\;.
\end{equation}

This metric represents a rotating black hole with cylindrical symmetry which will describe, via holography, a plasma that rotates with the same angular velocity $ \omega $ around a cylinder with radius $l$. 

For consistency, it is important to check if the metric \eqref{RotatingBHmetric} is a solution of the same Einstein equation satisfied by the AdS Black hole metrics metrics \eqref{BHAdS} and \eqref{CBH}. For both metrics the equation reads
\begin{equation} \label{Einstein}
R_{mn}  - \frac{1}{2} R \, g_{mn} + \Lambda \, g_{mn} = 0,
\end{equation}
where $R = -20/L^2$ ,  $\Lambda$ is the negative cosmological  constant and there is no energy density. Eq. \eqref{Einstein}  implies that the Ricci tensor $ R_{mn} $ is proportional  to the metric. Both $ R_{mn} $  and $ g_{mn} $ are second rank tensors that transform under coordinate transformations as
\begin{equation}
    T^{\prime}_{\alpha \beta}  = \frac{\partial x^m}{\partial x^{\prime \alpha }} \frac{\partial x^n}{\partial x^{\prime \beta }} T_{m n }\,\, 
    \equiv \, M^{m n }_{\alpha\beta} \,  T_{m n } . \label{trans}
\end{equation}
The transformation matrix $ M^{m n }_{\alpha\beta} $ for the present case is easily obtained from the inverse of the coordinate transformations \eqref{T1} and \eqref{T2},
\begin{eqnarray}
	t &=& \frac{1}{\sqrt{1-l^2 \omega^2}} \left(t^\prime -  l^2 \omega \phi^\prime \right)\;, \\
	\phi & = & \frac{1}{\sqrt{1-l^2 \omega^2}} \left(\phi^\prime -   \omega t^\prime \right) \;.\end{eqnarray}

The curvature scalar is not affected by coordinate transformations. Contracting  $ M^{m n}_{\alpha\beta} $ with eq. \eqref{Einstein} and using \eqref{trans} one finds
\begin{equation} \label{Einstein2}
R^{\prime}_{\alpha\beta}  - \frac{1}{2} R \, g^{\prime}_{\alpha\beta} + \Lambda \, g^{\prime}_{\alpha\beta} = 0,
\end{equation}
so,  the metric \eqref{RotatingBHmetric} satisfies the same Einstein equation satisfied by the original metric \eqref{CBH}.

The Lorentz boost affects the relation between the Hawking temperature $ T$ of the black hole and the horizon position. Rewriting the rotating metric \eqref{RotatingBH}  in the canonical form, see \cite{LEMOS199546, BravoGaete:2017dso}, one obtains
\begin{eqnarray}\label{canon}
	ds^2 &=& -N(z) dt^2 + \frac{L^2}{z^2}\frac{dz^2}{f(z)} + R(z)\left( d\phi + P(z) dt\right)^2  \cr && + \frac{L^2}{z^2} \sum_{i-1}^2 dx_i^2\;,
\end{eqnarray}
with 
\begin{eqnarray}
	N(z) &=& \frac{L^2}{z^2} \frac{ f(z) (1-\omega^2 l^2)}{1- f(z)\omega^2 l^2}\;, \\
	R(z) &=& \frac{L^2}{z^2}\left( \gamma^2 l^2 -  f(z) \gamma^2 \omega^2 l^4\right)\;, \\
	P(z) &=& \frac{\omega(1-f(z))}{1- f(z)\omega^2 l^2}\;. 
\end{eqnarray}

Now, defining $h_{00} = - N(z)$, the   temperature can be obtained from the surface gravity formula \cite{Zhou:2021sdy}:
\begin{eqnarray}\label{HT}
	T &=& \vert \frac{\kappa_G}{2\pi} \vert = \bigg\vert \frac{\lim_{z\rightarrow z_h}- \frac{1}{2} \sqrt{\frac{g^{zz}}{-h_{00}(z)}}h_{00,z}}{2\pi}\bigg \vert = \frac{1}{\pi z_h} \sqrt{1-\omega^2 l^2}\;, 
\end{eqnarray}
where $\kappa_G$ is the surface gravity, and $g^{zz}$ the $z-z$ component of the inverse of the cylindrical black hole metric.

The standard entropy of the black-hole is determined by $S=\frac{1}{4}A$, where $A$ is the area of the event horizon. In our case, it takes the form
\begin{equation}
s=\frac{2 \pi L^3}{\kappa^2}\frac{1}{z^{3}_{h}\sqrt{1-w^2l^2}}\;,
\end{equation}
where $s=S/V_{3D}$ is the  entropy density. For the mass and angular momentum, the result has already been determined\footnote{ The authors use the method developed by Regge-Teitelboim \cite{Regge:1974zd,Banados:1992gq}. The approach consists of writing the gravity action into the Hamiltonian form where a surface term is introduced to ensure that the  Hamiltonian equation is satisfied. Then, the  task is to find the surface term and associate it with the mass and angular momentum. For more details see refs.\cite{Regge:1974zd,Banados:1992gq}.} in the refs. \cite{PhysRevD.97.024034,BravoGaete:2017dso}. For this reason, we only present the result found, given by

\begin{eqnarray}
M&=&\frac{L^3}{\kappa^2}\frac{\omega^2l^2+3}{2(1-\omega^2l^2)}\;,
\cr \\
J&=&\frac{2L^3}{\kappa^2}\frac{\omega l}{z_h^{4}(1-w^2l^2)}\;,
\end{eqnarray}
where $M=m/V_{3D}$ and $J=j/V_{3D}$ are the mass and angular momentum densities, respectively. One can check that the first law of thermodynamic $dM= Tds + \omega dJ$ is satisfied for the rotating black hole geometry, as discussed in \cite{PhysRevD.97.024034,BravoGaete:2017dso}. In the Appendix A, we perform the calculation of the thermodynamic quantities for the  hard wall model, using the holographic prescription, finding the same mass, angular momentum and entropy.

\section{Regularized free energy density in hard and soft wall models} 

The hard \cite{Polchinski:2001tt,Boschi-Filho:2002wdj,Boschi-Filho:2002xih} and soft  \cite{Karch:2006pv} wall holographic AdS/QCD models consist in introducing an energy parameter in the AdS geometry. This parameter is interpreted, in the gauge theory side of the gauge/gravity duality, as an infrared (IR) cutoff. In the first case, this is done by just imposing that the $z$ coordinate of the AdS spaces has a maximum value: $ 0 \le z \le z_0$. In the second case one introduces in the geometry a dilaton background $ \Phi (z)  $ that contains an energy parameter. 

One can write the five-dimensional  gravitational action for both models, at zero temperature,  in the general form\cite{Herzog:2006ra,BallonBayona:2007vp}
\begin{equation}\label{action1}
	I = - \frac{1}{ 2 \kappa^2} \int_0^{z_0} dz\int d^4x \sqrt{g} e^{-\Phi}\left( R - \Lambda \right) \; =  \frac{4}{L^2\kappa^2} \int_0^{z_0}\int d^4x \sqrt{g} e^{-cz^2},     
\end{equation}
where $\kappa$ is the gravitational coupling associated with the Newton constant
and, as mentioned in the previous section, the cosmological constant and the curvature are related to the AdS radius by  $\Lambda = \frac{3}{5} R = \frac{-12}{L^2}$. 
The hard wall model corresponds to choosing $ c = 0$ and taking $ 1/z_0 $ as the IR energy parameter. For the soft wall $z_0 \to \infty $ and $ \sqrt{c} $ is the energy parameter.

In order to perform the  analysis of the Hawking-Page transition for the cylindrical rotating AdS spaces,  we consider the metric \eqref{RotatingBHmetric} for the BH case and  a similar expression,  but with $ f(z) = 1 $, for the thermal AdS. In both cases, the determinant of the metric is $g = \frac{L^{10}}{z^{10}}$, such that the finite temperature version of \eqref{action1} reads
\begin{equation}
	I_{on-shell} = \frac{4L^3}{\kappa^2} \int_0^{z_{min}} dz\int d^4x z^{-5}e^{-cz^2}\;\,,
\end{equation}   
where $z_{min}$ is the minimum of $( z_0 , z_h )$. Note that for the thermal AdS there is no horizon, or equivalently $ z_h \to \infty. $
The integration over the spatial bulk coordinates $x$ is trivial, so we define an  action density dividing the on-shell action by the spacial volume factor of the bulk, $\mathcal{E} = \frac{1}{V_{3D}} I_{on-shell}$. For a compact time direction in the Euclidean signature, we have $ 0 \leq t < \bar\beta $,
where, as explained in the previous section, for the black hole geometry  $ \bar\beta = \beta = 1/T   $, while for the thermal AdS $ \bar\beta = \beta^\prime =   \sqrt{f(\epsilon)} / T $. Then, the action densities, including the ultraviolet regulator   $\varepsilon$ have the general form:
\begin{equation}
	\mathcal{E}_s(\varepsilon) = \frac{4L^3}{\kappa^2} \int_0^{\bar\beta}dt \int_\varepsilon^{z_f}dz\, z^{-5}e^{-cz^2}  \;.  
\end{equation}
For both the thermal AdS and the rotating AdS black hole the free energy densities are infinite in the limit $\varepsilon \rightarrow 0$. 

In order to obtain a finite quantity, one defines the regularized free energy density of the rotating black hole as the difference between the energy densities of the two geometries,  \begin{equation}\label{DeltaE}
	\bigtriangleup \mathcal{E}(\varepsilon) = \lim_{\varepsilon \rightarrow 0} \left[\mathcal{E}_{BH}(\varepsilon) - \mathcal{E}_{AdS}(\varepsilon) \right]\;, 
\end{equation}
where
\begin{eqnarray}
	\mathcal{E}_{BH}(\varepsilon) &=& \frac{4L^3}{\kappa^2} \beta \int_\varepsilon^{\text{min}(z_h,z_0)}dz\, z^{-5}e^{-cz^2}  \;, \label{DeltaEBH}\\
	\mathcal{E}_{AdS}(\varepsilon) &=& \frac{4L^3}{\kappa^2} \beta^\prime \int_\varepsilon^{z_0}dz\, z^{-5}e^{-cz^2}  \;. \label{DeltaEAdS}
\end{eqnarray}

The regularized free energy density \eqref{DeltaE} determines the Hawking-Page transition, as it measures the stability of the rotating black hole. When  $\bigtriangleup \mathcal{E}$ is positive (negative), the BH is unstable (stable), since the free energy density of the AdS space is smaller (greater) than the black hole one. From the gauge/gravity duality, this analysis concerning black hole (in)stability  corresponds to the   transition between the confined hadronic phase and the deconfined plasma phase.

\section{Confinement/Deconfinement transition in a rotating plasma}

The phase transition occurs when the regularized free energy density vanishes. The computation of equation \eqref{DeltaE} depends on the holographic model, \textit{i.e.}, on how the IR energy scale is introduced. Let us consider separately  the hard wall\cite{Polchinski:2001tt,Boschi-Filho:2002wdj,Boschi-Filho:2002xih}  and soft wall  \cite{Karch:2006pv} AdS/QCD models.

\subsection{Hard wall model}

In the hard wall model there is no dilaton field in the background, so we take $c=0$ in eqs.  \eqref{DeltaEBH} and \eqref{DeltaEAdS} with metrics \eqref{RotatingBHmetric} and  \eqref{cthermalAdS} respectively. The regularized free energy density  of eq. \eqref{DeltaE} then takes the form
\begin{equation}
	\centering
	\bigtriangleup \mathcal{E}(\omega l ,T) = \left\{\begin{aligned}
		\frac{L^3\pi}{\kappa^2\sqrt{1-\omega^2 l^2}} \frac{1}{2z_h^3}\;, \quad  z_0&< z_h\;, \\
		\frac{L^3\pi z_h}{\kappa^2\sqrt{1-\omega^2 l^2}}\left( \frac{1}{z_0^4}- \frac{1}{2 z_h^4} \right) \;, \quad z_0&> z_h \;. \\ 
	\end{aligned}\right.\label{FreeHW}
\end{equation}

Using the Hawking temperature \eqref{HT} for the rotating BH geometry, one automatically concludes that the Hawking-Page transition occurs at the critical temperature
\begin{equation}
	T_c^{(HW)}(\omega l ) = \frac{ 2^{1/4} \sqrt{1-\omega^2l^2}}{\pi z_0}\;=  
	\frac{ T_c^{(HW)} (0)} { \gamma(\omega l )}, 
	\label{TcHW}
\end{equation}
 where $ \gamma(\omega l )$ is the usual Lorentz factor and $T_c(0) $ is the critical temperature when the medium is not rotating. This result shows that the critical temperature depends on the QGP angular velocity. As $\omega l  $ increases, $T_c$ decreases.

\subsection{Soft wall AdS/QCD model}

In the soft wall AdS/QCD model \cite{Karch:2006pv} $ z_0 \to \infty $, while $c \neq 0$ is the parameter responsible for introducing the mass scale $\sqrt{c}$ in the model.   Equations \eqref{DeltaEBH} and \eqref{DeltaEAdS} now lead to   
\begin{equation}
	\bigtriangleup \mathcal{E}(\omega l , z_h ) \, = \,\frac{\gamma (wl)  L^3 \pi }{\kappa^2  z_h^3 } \bigg[ - e^{-c z_h^2} ( - 1 + c z_h^2 ) + \frac{1}{2} + 
	c^2 z_h^4 \text{Ei}\left( -c z_h^2 \right) \bigg] \,.\label{FreeSW}
\end{equation}
This is just the factor $ \gamma (w l)  $ times the free energy of the soft wall model without rotation obtained in Ref. \cite{Herzog:2006ra}. So, the condition $   \bigtriangleup\mathcal{E}(\omega l , z_h ) = 0 $ leads to the same relation found in this reference for the horizon position at the critical temperature:

\begin{equation}
    c z_h^2 = 0.419035 \,.
    \label{zhcritical}
\end{equation}
Defining a dimensionless temperature 
\begin{equation}
\label{dimlessTCSoft}
    \bar{T} = T/\sqrt{c} 
\end{equation}
 and using the relation between the horizon position and the temperature for the case with rotation, given by eq. \eqref{HT}, one finds:
\begin{equation}
\label{Tcsoft}
   \bar{T}_c (\omega l)  = \frac{ 0.491728}{  \gamma (\omega l )}  = \frac{ \bar{T}_c (0) } {\gamma (\omega l )} \,,
\end{equation}
that is the same dependence on $\omega l $ found in the case of the hard wall model. 
In  Fig. 1, we plot the free energy densities of the rotating plasma, in the soft wall case,  as a function of $\bar{T}$, at fixed values of $\omega l $. As one can see, the critical temperatures, defined by eq. \eqref{Tcsoft}, decreases as the rotational velocity  increases, in the same way as in the hard wall model. This result does not depend on the choice of the infrared parameter $c$, since we are working with the dimensionless temperature \eqref{dimlessTCSoft}.

\begin{figure}[!htb]
	\centering
	\includegraphics[scale=0.55]{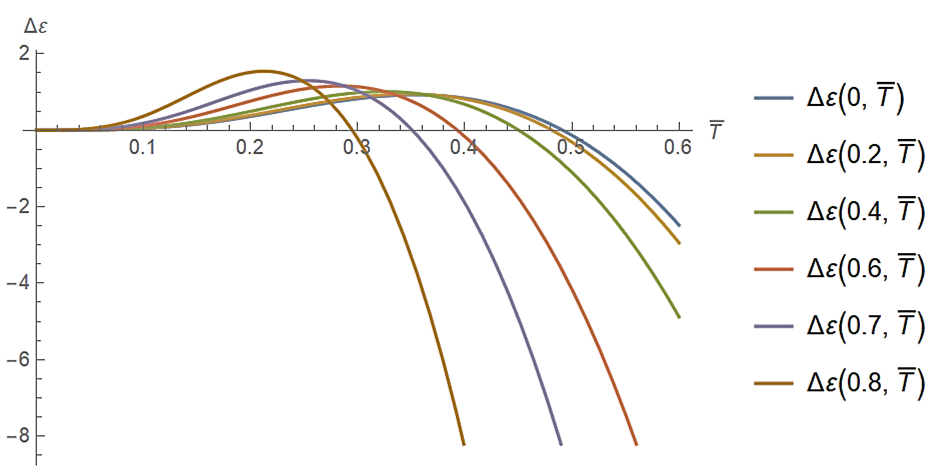}
	\caption{Free energy density of rotating plasma ($\bigtriangleup \mathcal{E}$) as a function of $\bar{T}$, at different rotational velocities $\omega l $, for the soft wall model. }
\end{figure}

In the Fig. 2 (A), we plot the free energy density of the plasma in the soft wall case,  as a function of $\bar{T}$ and $\omega l $.  The curve $\bar{T}_c(\omega l )$ is determined by the intersection between $\bigtriangleup \mathcal{E}(\omega l ,\bar{T})$ and the plane $\bigtriangleup \mathcal{E}(\omega l ,\bar{T}) = 0$, indicated in the Fig. 2 (B).  This intersection, for $ 0 \leq \omega l \leq 1$, corresponds to the blue curve plotted in Fig. 3. In the same figure, the orange curve represents ${\bar T}_c(\omega l)$ of the hard wall case.

\begin{figure}[!htb]
	\centering
	\includegraphics[scale=0.52]{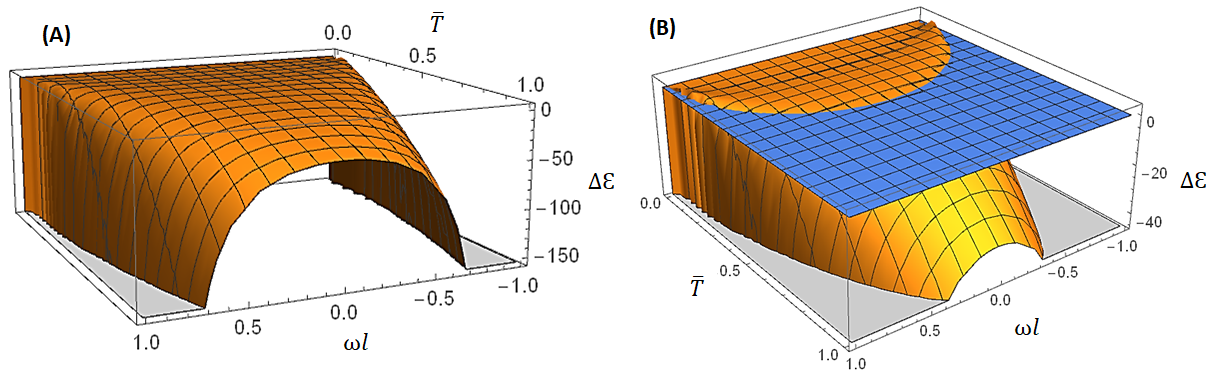}
	\caption{(A) Free energy density of rotating plasma as a function of $T$ and $\omega l$; (B) Hawking-Page curve for critical temperatures represented by the intersection between $\bigtriangleup \mathcal{E}(\omega l,\bar{T})$ and the plane $\bigtriangleup \mathcal{E}(\omega l,\bar{T}) = 0$.  }
\end{figure}

\begin{figure}[!htb]\label{Tcvswl}
	\centering
	\includegraphics[scale=0.65]{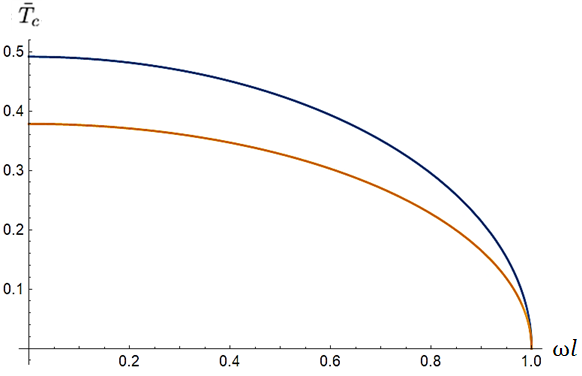}
	\caption{ Dimensionless  critical temperature $\bar{T}_c$ versus rotational velocity ($\omega l$)  in the soft wall (blue), and hard wall (orange) models. For the hard wall we use: $\bar{T}_c \equiv  z_0  \,  T_c $}
\end{figure}

\subsection{Comparison of the results for the two models}

The holographic prediction for both  hard wall and soft wall models is that $T_c$ decreases as the absolute value of the  velocity $\omega l $ increases. So, for a rotating plasma the confinement/deconfinement phase transition occurs at lower temperatures than for a static plasma.  The fact that the critical temperature has the very same dependence on the velocity for both models, as shown in equations \eqref{TcHW} and \eqref{Tcsoft}, was not expected since these models are very different and produce, for example,  different mass spectra for hadrons. The explanation is that the free energies of the hard and soft wall models, given respectively by equations \eqref{FreeHW}  and \eqref{FreeSW}, are equal to the free energies without rotation times the factor $ \gamma (w l)$. In other words, rotation does not affects the  horizon position corresponding to the condition of vanishing free energy. Therefore, equation \eqref{HT} relating the horizon position to the temperature implies the same dependence on  $\omega l $  found here for the critical temperature of both models.

\section{Analysis of the results and conclusions}

In noncentral relativistic heavy ion collisions, one can estimate the angular velocity and the size (or radius) of the rotating QGP that is produced -- see, for instance, \cite{STAR:2017ckg, Jiang:2016woz, Sievert:2019zjr, Braguta:2021jgn}. In \cite{STAR:2017ckg}, the authors argued that experimental results for $\Lambda$, $\Lambda^-$ baryons polarization give the average value $\omega \sim 6 \;\text{MeV}$ for the angular velocity of the plasma. Hydrodynamic simulations of heavy-ion collisions worked out in \cite{Jiang:2016woz} predict even larger magnitudes of the angular velocity with $\omega  \sim 20-40\; \text{MeV}$. Meanwhile, typical values for the QGP size were estimated in \cite{Sievert:2019zjr}, in which the authors used relativistic hydrodynamics to make predictions for a possible future run of ArAr and OO collisions at the Large Hadron Collider.

With these information, one can use equations \eqref{TcHW} and (or) \eqref{Tcsoft}  to compute some typical value of the variation of the critical temperature $ \bar{T}_c (wl) /  \bar{T}_c (0)  $ caused by rotation of the plasma. For example, for  radius $ l = 5\, \text{fm} $ and $ \omega = 20 $ MeV one finds
\begin{equation}
    \bar{T}_c (\omega l) /  \bar{T}_c (0)  = \frac{1}{\gamma ( 0.5) } = 0.87 
\end{equation}

The result that the  critical temperature of the confinement/deconfinement transition  decreases with  increasing angular velocity is in agreement with recent studies of  this transition in rotating QCD using phenomenological models \cite{Chen:2020ath,Chernodub:2020qah,Fujimoto:2021xix}.  The authors of Ref. \cite{Chen:2020ath} investigated the rotation effect on the deconfinement phase transition in an Einstein-Maxwell-Dilaton(EMD) model in pure gluon and two-flavor holographic QCD models in the presence of a chemical potential.
The authors extended a model proposed in Ref. \cite{Dudal:2017max} to finite angular velocity. Then, from the analysis of thermodynamic quantities,  it is shown that the critical temperature   decreases with angular velocity. It is also shown
that the pressure, energy density, entropy density and specific heat are enhanced by the angular velocity. To get full understanding of the phase transition, and to check the physical contents of the geometric phase transition, the authors of \cite{Chen:2020ath} investigated the order parameter of the deconfinement transition, i.e. the loop operators. In contrast, here we analyzed the rotation effect considering the Hawking/Page phase transition in the hard wall model and soft wall model without  chemical potential. The analytical expressions (\ref{TcHW}) and (\ref{Tcsoft}) show that the critical temperature decrease with the increase of the angular velocity which is in qualitative agreement with the result found \cite{Chen:2020ath}. 
The same behaviour was found also in Nambu-Jona-Lansinio model \cite{Jiang:2016wvv}, in which  the critical temperature in rotating QCD decreases due to suppression of the chiral condensate.

On the other hand, the simulations of relativistic rotation  on the confinement/deconfinement  phase transition in gluodynamics  lattice \cite{Braguta:2021jgn}, using different types of boundary conditions (Open, Periodic and Dirichlet), found that the critical temperature increases with increasing angular velocity for all boundary conditions and all lattice parameters used in the calculations, showing up an opposite behaviour if compared to our result and to the other models \cite{Chen:2020ath,Chernodub:2020qah,Fujimoto:2021xix, Jiang:2016wvv} cited above. However, the same authors of \cite{Braguta:2021jgn} have recently presented the first lattice results in \cite{Braguta:2021ucr} (preliminary) for the confinement/deconfinement phase transition taking into account dynamical fermions. Such a preliminary result shows  that the critical temperature with  rotating fermions decreases with the grow of angular velocity, thus also in agreement with our prediction.
For previous studies of the thermodynamics of rotating black holes see \cite{Hawking:1998kw,Hawking:1999dp,Gibbons:2004ai} and for rotating quark gluon plasma in the context of holography, see for example  \cite{Miranda:2014vaa,Mamani:2018qzl,Arefeva:2020jvo,Golubtsova:2021agl}.

\appendix
 \section{Holographic renormalization and thermodynamics}
 
In section 2 we have presented the thermodynamic quantities of the rotating cylindrical black hole obtained from general relativity. Now, we will study the thermodynamics of rotating black holes in the hard-wall model.  We will not perform this analysis for the soft wall  model because in this case we cannot compare the results with the general relativity ones, since this phenomenological model is not a solution to Einstein equation. We will consider the hard wall model with $ z_o > z_h $, when the Einstein equations are satisfied, using the holography prescription. This can be done by evaluating the Euclidean bulk on-shell action together with the appropriate counterterm and the Gibbons-Hawking action. The counterterm is necessary  to handle with the UV divergence present in the bulk and Gibbons-Hawking actions, as prescribed by the holographic renormalization. For this reason, we consider the following gravitational action

\begin{eqnarray}\label{total}
I_{T}&=& I_{bulk}+ I_{GH}+ I_{c.t.}\,.
\end{eqnarray}
The on-shell bulk action has already been calculated (\ref{action1}) and reads
\begin{equation}\label{bulk}
I_{bulk}=\frac{L^3V_{3D}}{\kappa^2}\beta\left( \frac{1}{\epsilon^4}-\frac{1}{z^{4}_{h}}\right)\,,
\end{equation}
where $\epsilon$ is the ultraviolet regulator. The Gibbons-Hawking action is calculated from
\begin{equation}\label{GravityActionBH}
I_{GH}=-\frac{1}{\kappa^2}\int d^4x \sqrt{h}K\,,
\end{equation}
where $K$ is the trace of the extrinsic curvature at a boundary hypersurface,
\begin{equation}
K=\nabla_{a}n^a=\frac{1}{\sqrt{g}}\partial_{a}\left( \sqrt{g} n^{a} \right)\;,
\end{equation}
with $n^a \, = \, \left( -z\sqrt{f(z)}/L,0,0,0,0 \right)$ being an unitary vector normal to the boundary, and $h$ the determinant of the induced metric $h_{\mu\nu}$ at the boundary, such that  $\sqrt{h}=\frac{L^4\sqrt{f(z)}}{z^4}$. Therefore, the surface action  for the cylindrical rotating black hole reads
\begin{equation}\label{GH}
I_{GH}=- \frac{4L^3V_{3D}}{\kappa^2}\beta\left(\frac{1}{\epsilon^4}-\frac{1}{2 z_{h}^4} \right)\,.
\end{equation}
The counterterm action for the $AdS_5$  can be written as
\begin{equation}
S_{ct}=\frac{1}{\kappa^2} \int d^4x \sqrt{h}\frac{3}{L}\;,
\end{equation}
then, replacing $\sqrt{h}$, one finds 
\begin{equation}\label{ct}
I_{ct}= \frac{3L^3V_{3D}}{\kappa^2}\frac{\beta}{\epsilon^4}\sqrt{f(\epsilon)}\simeq  \frac{3L^3V_{3D}}{\kappa^2} \beta \left(\frac{1}{\epsilon^4}-\frac{1}{2 z^4_{h}} \right)\;.
\end{equation}
Now, replacing (\ref{bulk}), (\ref{GH}) and (\ref{ct}) into (\ref{total}), one gets the total action density
\begin{equation}
\mathcal{E}_{T}=-\frac{L^3}{\kappa^2}\frac{\beta}{2 z_{h}^{4}}=-\frac{L^3 \pi^4}{2\kappa^2}\frac{T^3}{(1-w^2l^2)^2}=-\frac{L^3 \pi^4}{2\kappa^2}\frac{1}{\beta^3(1-w^2l^2)^2}\,.
\end{equation}
Using the semi-classical approximation, the on-shell action is related to the partition function $Z$ and the thermodynamic potential $\Phi$ via
\begin{equation}
Z=e^{-\mathcal{E}_T}\;, \quad \text{with} \quad  \mathcal{E}_T=\beta \Phi\;,
\end{equation}
from which the potential becomes
\begin{equation}
\Phi=-\frac{L^3 \pi^4}{2\kappa^2}\frac{T^4}{(1-w^2l^2)^2}=-\frac{L^3 \pi^4}{2\kappa^2}\frac{1}{\beta^4(1-w^2l^2)^2}\,.
\end{equation}
In the end, the thermodynamic quantities are derived from $\Phi$ using  the thermodynamics relations:
\begin{eqnarray}
&s&=-\frac{\partial \Phi}{\partial T}=\frac{2L^3 \pi^4}{\kappa^2}\frac{T^3}{(1-w^2l^2)^2}=\frac{2L^3 \pi}{\kappa^2}\frac{1}{z_{h}^{3}\sqrt{1-w^2l^2}}\;, \cr &J&=-\frac{\partial \Phi}{\partial \omega}=\frac{2L^3 }{\kappa^2}\frac{\omega}{z_{h}^{4}(1-w^2l^2)}\;, \cr &M&=\Phi +sT+ \omega J =\frac{L^3 }{2\kappa^2}\frac{(\omega^2+3)}{z_{h}^{4}(1-\omega^2l^2)}\;.
\end{eqnarray}
As we can see, the first law $dM= Tds + \omega dJ$ is automatically satisfied, in agreement with the result found in \cite{PhysRevD.97.024034,BravoGaete:2017dso}.

\noindent \textbf{Acknowledgments}: The authors are supported by  CNPq - Conselho Nacional de Desenvolvimento Cient\'ifico e Tecnol\'ogico. This work received also support from  Coordena\c c\~ao de Aperfei\c coamento de Pessoal de N\'ivel Superior - Brasil (CAPES) - Finance Code 001. 

\bibliographystyle{utphys2}
\bibliography{library}

\end{document}